\documentclass[twoside,twocolumn]{article}

\usepackage[sc]{mathpazo} 
\usepackage[T1]{fontenc} 
\usepackage{microtype} 
\usepackage{graphicx} 

\usepackage[english]{babel} 

\usepackage[hmarginratio=1:1,top=32mm,columnsep=20pt]{geometry} 
\usepackage[hang, small,labelfont=bf,up,textfont=it,up]{caption} 
\usepackage{booktabs} 

\usepackage{lettrine} 

\usepackage{enumitem} 
\setlist[itemize]{noitemsep} 

\usepackage{abstract} 

\usepackage{titlesec} 
\renewcommand\thesection{\Roman{section}} 
\renewcommand\thesubsection{\roman{subsection}} 
\titleformat{\section}[block]{\large\scshape\centering}{\thesection.}{1em}{} 
\titleformat{\subsection}[block]{\large}{\thesubsection.}{1em}{} 

\usepackage{fancyhdr} 
\usepackage{titling} 

\usepackage{hyperref} 

\begin{titlepage}
\pretitle{\Huge\bfseries} 
\posttitle{} 
\title{On the Possibility of Discovering Exoplanets within our Solar System} 
\end{titlepage}
\author{%
\textsc{J. A. Paice} \\[1ex] 
\normalsize Department of Physics and Astronomy, University of Manchester, \\ \normalsize Oxford Rd, Manchester, M13 9PL, UK\\ 
\normalsize \href{mailto:john_paice@hotmail.co.uk}{john\_paice@hotmail.co.uk} 
\and 
\textsc{J. J|C. Watkins} \\[1ex] 
\normalsize Department of Unexpected Geography, Unseen University, \\ \normalsize Sator Square, Ankh-Morpork, Ankh-Morpork \\ 
}
\date{1 April 2022} 
\renewcommand{\maketitlehookd}{%
\begin{abstract}
\noindent It has previously been suggested that the rate of exoplanet discovery, doubling roughly every 39 months, is indicative of a runaway increase in the number of exoplanets in our galaxy. In this paper, we posit that - due to the finite nature of space in the milky way - it will become increasingly likely that one of these exoplanets will be found within our solar system. We calculate the odds of this occurring pass 50\% on Friday 9th December 2146. We go on to suggest novel methods for influencing where this exoplanet may be discovered, note possible drawbacks of the discovery, and finally explore how the previously-hypothesized `exoplanet singularity' (both figurative and literal) could be averted.
\end{abstract}
}

\begin{document}
\maketitle

\section{Introduction} \label{sec:Intro}

The discovery of new planets has long been a goal of astronomical research. The field of exoplanets has currently found 4,940 examples of such bodies\footnote{\href{https://exoplanetarchive.ipac.caltech.edu/}{https://exoplanetarchive.ipac.caltech.edu/}}. However, while each new exoplanet discovery is undoubtedly a great achievement, \textit{in situ} studies of planets around other stars are currently outside our abilities.

However, if an exoplanet were to be found closer, such as in orbit around the Sun, it would be a much needed boon for the field. A literature search for planets in our solar system has revealed between eight or nine currently-known examples\footnote{We exclude dwarf planets from this count, because they are small and no-one likes them.}, with one tantalising discovery within the sun's habitable zone
\cite{Yahweh_Bible_4004, Munroe_1371}. However, in recent years, these discoveries have tailed off; in fact, no new planets have been found to orbit the sun within the last hundred years, despite numerous searches (For a review, see \cite{OMalleyRosset_PlanetX_2020}). Using the methods of popular geography, this means that we are overdue for such an event.

The discovery of new exoplanets has recently been noted as following a process known as Mamajek's Law: the number of known exoplanets doubles every 39 months. A recent paper \cite{Lund_ExistentialThreat_2021} (hereafter L21) notes a minor concern with this theory when taken to its logical extreme - i.e., the collapse of our galaxy into an all-consuming black hole.

From this theory, we pose a possible upside. As the number of new exoplanets increases, following the distribution laid out by Mamajek's law, there will come a point when an exoplanet must be discovered inside our solar system. Naturally, this will be of great interest to all scientists in the field of exoplanets (hereafter `exoplaneteers').

In this paper, we will investigate when this discovery is most likely to occur, consider possible additional pitfalls arising from L21, and also pose several novel methods for directing these discoveries.

\vspace{5mm}

\textit{A note on terminology:} In this paper we use the terms `planet' and `exoplanet' interchangeably. Since we do not live in an anthropocentric universe, we would not dream of thinking ourselves so self-important as to assume planets are different from exoplanets. After all, all planets are exoplanets - from a certain point of view \cite{Kenobi_4ABY}.

\section{Methods} \label{sec:Method}

A key premise from L21 was that the number of discovered exoplanets doubled every 39 months. The author went on to deduce thus; the total mass of these exoplanets would mean that the mass of the Milky Way, contained within a 10\,kpc radius, would exceed that of the corresponding Schwarzchild mass in 2252.

However, we can make approximations to deduce when it is likely that one of these exoplanets would be discovered within the solar system. We postulate a simple model of the Milky Way as a 2-dimensional plane, emitting exoplanets in all directions\footnote{\href{https://en.wikipedia.org/wiki/Spherical_cow}{https://en.wikipedia.org/wiki/Spherical\_cow}}, and then find the area of the Solar System compared to the total area of the 2D plane contained within its orbit:

\begin{equation}
P = \frac{A_{SS}}{A_{MW}} = \frac{R_{SS}^2}{R_{MW}^2}
\label{eqn:probability}
\end{equation}

where P is the possibility of any given new exoplanet within the Milky Way (MW) being within the boundaries of the Solar System (SS), A is the area, and R is the radius.

We take the outer edge of the solar system as 4.5 $\times$ 10$^9$\,km (Google 2021), this being the orbit of Sol-i (known colloquially as 'Neptune'), the furthest of the Milky Way's exoplanet population; this marks the rough boundary that can be reached by our current technology. For the radius of the Milky Way, we follow the example of L21, noting that all exoplanets are found within 10\,kpc. From these figures, we find that any given exoplanet has a 1 in 4.7 $\times$ 10$^{15}$ chance of being found within our solar system.

We thus assume that once that many exoplanets are discovered, it becomes likely that one of those will be within the solar system. We note that this is just an approximation; more advanced statistical modelling, perhaps utilising the curious phenomenon of most exoplanets being found to congregate around the solar system \cite{Huff_1954}, is beyond the scope of this paper.

L21 gives the following equation for Mamajek's law of the number of exoplanets:

\begin{equation}
N_{planets} = 10.25*1.24^{\left(year - 1990\right)}
\label{eqn:mamajeks_law}
\end{equation}

Reversing this equation tells us that we will reach 4.7 $\times$ 10$^{15}$ exoplanets - and thus an exoplanet in our solar system will be discovered - on Friday 9th December 2146, just after 22:36 UTC\footnote{Once again fitting the tradition that all big discoveries must disrupt an astronomer's weekend.}.


\section{Discussion} \label{sec:Discuss}

While an exoplanet being discovered inside our solar system will have great potential for scientific discovery, it would be beneficial if this planet were to be found close to Earth for logistical reasons. Fortunately, the history of exoplanet research provides ample anecdotal evidence to aid in this fact.\footnote{We realise that anecdotal evidence alone is not sufficient to claim a discovery. Therefore, we took a lot of anecdotes, and averaged them out. This is called science and is perfectly acceptable.}.

An extensive literature search reveals that the vast majority of papers regarding exoplanet discoveries are about detections of new exoplanets - conversely, there are relatively few papers about non-detections of exoplanets. \textit{Ipso facto}, the vast majority of exoplanet searches lead to a discovery of exoplanets. We thus postulate that exoplanets will most likely be found wherever exoplaneteers are observing (especially considering that there are currently no known exoplanets in locations that have not been observed). This targeted approach to exoplanet discovery builds on previous proposals in the field \cite{Lund_2016}.

From this, we arrive at a logical conclusion: The more exoplaneteers we have observing a location, the more likely it is that an exoplanet will be discovered there.

The ramifications of this theory are multifarious. For instance, we can use simultaneous observations on multiple different telescopes to `triangulate' a discovery on a particular location. The best locations to target should be trivial to travel to, and ideally not upset the orbital motions of other Solar System bodies; this could include Lagrange points L3 and L4 in the Earth-Sun system\footnote{Other Lagrange points are available. An exoplanet discovered at L1, for example, would be easily detected through the transit method and may provide many benefits to observational astronomy. However, we stress that L2 in particular should \textit{not} be selected, as various telescopes that reside there - such as JWST - are not equipped for unexpected lithobraking.}.

However, we note that there are complications that may result from this theory of exoplaneteer idealism - that is, that exoplaneteers may inadvertently discover an exoplanet outside of the targeted locations. For example, it would be disadvantageous if an exoplanet were to be discovered inside an exoplaneteer's living room. We thus recommend that all exoplaneteers should be blindfolded when not directly observing Lagrange points.

We should also be aware of the nature of the exoplanet that may be discovered. Planets appearing in our solar system have been posited before, with less than stellar results \cite{PedlerDavis_1966}, and other authors from the same institute have noted possible events near to 2146 that may disrupt ongoing studies \cite{Subotsky_1966}. We also note a worry that every new exoplanet could also host its own exoplaneteers, which may accelerate Mamajek's law.

\vspace*{0.3cm}
\noindent
    
\section{Conclusions}

In this paper, we calculate the statistical likelihood that an exoplanet will be discovered in our solar system by the end of 2146. This will give us roughly 102 years to research it before the black hole posited by L21 is due to consume all matter in the Milky Way.

We deduce that many exoplanet discoveries are due to the observational abilities of exoplaneteers; i.e. new exoplanets are only found because they're looking for them\footnote{``It's probably because it's quantum" \cite{Pratchett_Pyramids_1989}.}.

Ergo, we pose a novel solution to Lund's black hole; when an exoplanet is discovered in our solar system, we send all exoplaneteers to it, where they will have ample opportunity to observe it \textit{in situ}. With this, they will hopefully no longer feel the need to find new exoplanets, thus finally breaking Mamajek's law.

\vspace*{0.3cm}
\noindent



\begin{figure}
    \includegraphics[width=\columnwidth]{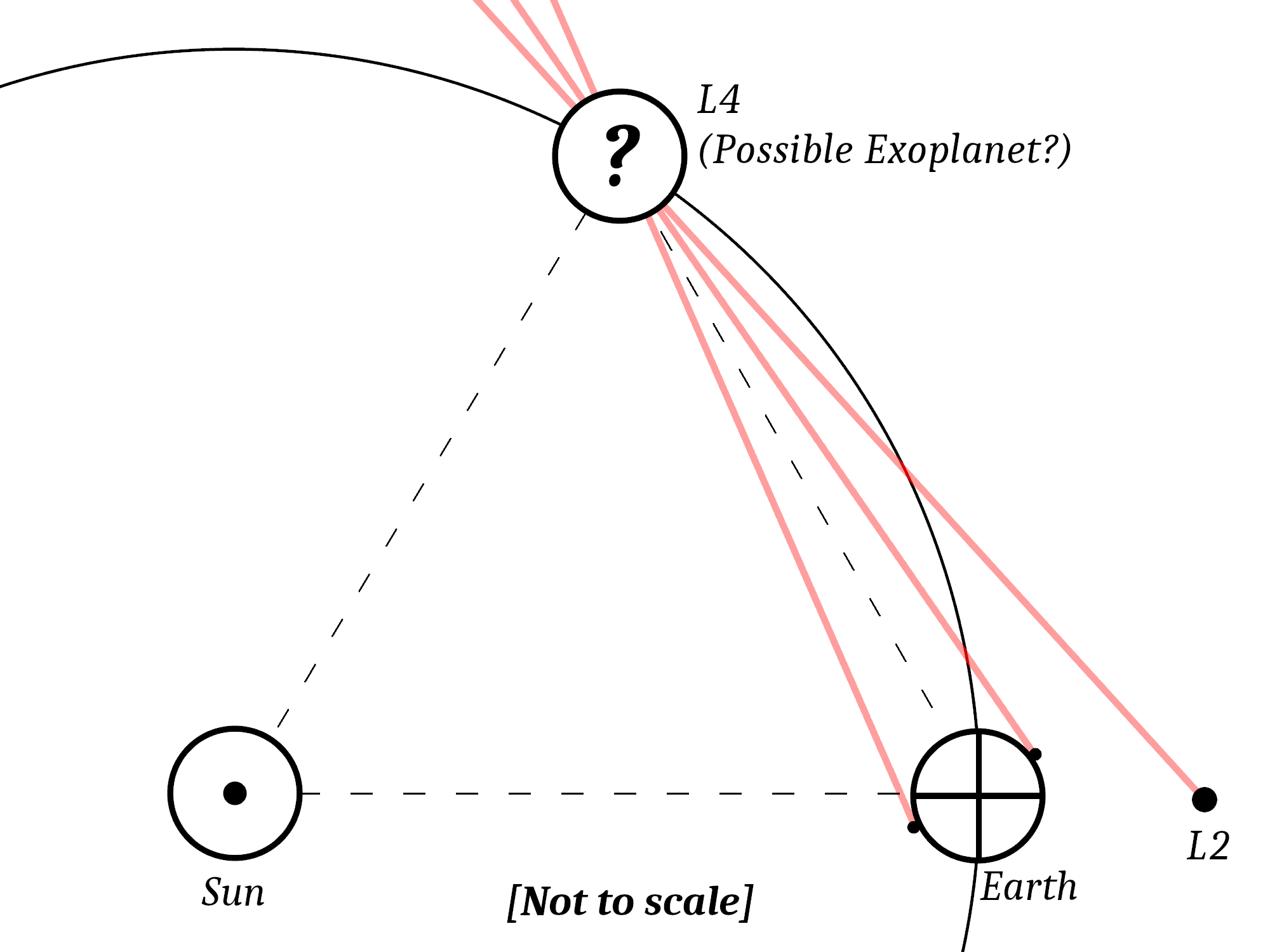}
    \caption{Possible discovery method for new exoplanets.}
    \label{fig:input_ps_lor}
\end{figure}

\section*{Acknowledgements}
	
The authors would like to thank Michael B. Lund for inspiring this paper. We thank the anonymous referee for not looking at this paper. GNU Terry Pratchett. This research has made use of NASA’s Astrophysics Data System.






	
	

	

	
	
	
	
	

\label{lastpage}
\end{document}